\documentclass[mathleft]{an}
\usepackage{graphicx,times}
\newcommand{\bfu}{{\mbox{\boldmath $\vec u$}}}
\newcommand{\bfk}{{\mbox{\boldmath $\vec k$}}}
\newcommand{\bfe}{{\mbox{\boldmath $\vec e$}}}
\newcommand{\bfB}{{\mbox{\boldmath $\vec B$}}}

\newcommand{\p}{\partial}
\newcommand{\Om}{{\it \Omega}}

\def\lsim{\lower.4ex\hbox{$\;\buildrel <\over{\scriptstyle\sim}\;$}} 
\def\rsim{\lower.4ex\hbox{$\;\buildrel >\over{\scriptstyle\sim}\;$}}
\def\beg{\begin{eqnarray}}
\def\ende{\end{eqnarray}}
\Yearpublication{2008}
\Yearsubmission{2008}
\Month{5}

\Volume{}
\Issue{}
\DOI{}

\begin{document}
\title{Helical magnetorotational instability  of Taylor-Couette flows 
in  the  Rayleigh limit and for  quasi-Kepler rotation}
\author{G. R\"udiger\thanks{Corresponding author: gruediger@aip.de} \and M. Schultz}
\titlerunning{MRI with helical fields}
\authorrunning{G. R\"udiger \& M. Schultz}
\institute{Astrophysikalisches Institut Potsdam, An der Sternwarte 16, D-14482 Potsdam, Germany
}
\received{2008 May 14} \accepted{2008 Jul 2} \publonline{2008} 

\keywords{turbulence -- magnetohydrodynamics -- magnetorotational instability}
\abstract{The magnetorotational instability (MRI)  of differential rotation  under the simultaneous 
presence of axial and azimuthal  components of the (current-free) magnetic field is considered. For rotation with uniform specific angular momentum the  MHD equations for 
axisymmetric  perturbations  are solved  in a local short-wave  approximation. 
All the solutions  are overstable for $B_z \cdot B_\phi\neq 0$ with eigenfrequencies   approaching   
 the viscous frequency. For more flat rotation laws the results of the local approximation do not comply with the results of a global calculation of the MHD instability of   Taylor-Couette flows between rotating cylinders. -- With  $B_\phi$  and  
 $B_z$  of the same order the traveling-mode solutions  are  also prefered for flat rotation laws such as the quasi-Kepler rotation.  For magnetic Prandtl number ${\rm Pm}\to 0$ they scale  with the   Reynolds number  of  rotation rather than with the magnetic Reynolds number (as for standard MRI) 
so that they  can easily be realized in MHD laboratory experiments.
-- Regarding the  nonaxisymmetric modes one  finds a remarkable influence of the ratio  $B_\phi/B_z$ only for the extrema. For $B_\phi\gg B_z$ and for not too small  Pm the nonaxisymmetric modes  dominate the traveling axisymmetric modes. For standard MRI  with $B_z\gg B_\phi$, however,   the critical Reynolds numbers of the nonaxisymmetric modes  exceed the values for the axisymmetric modes by many orders  so that they   are never  prefered.
}
\maketitle

%%%%%%%%%%%%%%%%%%%%%%%%%%%%%%%%%%%%%%%%%%%%%%%%%%%%%%%%%%%%%%%%%%%%%%%%%%%%%%%%%%%%%%%%%%%%%%
\section{Introduction}
%%%%%%%%%%%%%%%%%%%%%%%%%%%%%%%%%%%%%%%%%%%%%%%%%%%%%%%%%%%%%%%%%%%%%%%%%%%%%%%%%%%%%%%%%%%%%
It has been shown in previous publications starting with  Hollerbach \& R\"udiger (2005) that the magnetorotational instability (MRI)  under 
the presence of both current-free axial and azimuthal components of the magnetic field ('Helical' fields, hence HMRI) is always characterized by 
an eigenoscillation frequency. In 
combination with the vertical wavenumber the resulting instability pattern is thus an axisymmetric  wave  traveling along 
the rotation axis. In all our considerations the magnetic Prandtl number Pm plays the basic role. For $\rm Pm \to 0$ the HMRI scales with the Reynolds number Re rather than with the magnetic Reynolds number Rm as it does for  the standard magnetorotational instability   for  an axial magnetic field.   The questions arise whether this frequency reflects the geometry of the magnetic field and 
whether it is observable  with real astrophysical objects such as protoneutron stars and/or accretion disks. Also the relation to the  Azimuthal MagnetoRotational Instability (AMRI, see R\"udiger et al. 2007b) for (current-free) toroidal fields which is nonaxisymmetric and which scales with the magnetic Reynolds number  for $\rm Pm \to 0$ must be considered.  We find  the AMRI only weakly (if ever) influenced by the addition of an axial field which is not much stronger than the toroidal field.  For $\rm Pm\simeq 1$ the difference between Re and Rm disappears so that the main differences of the instabilities also disappear.

In the present paper we start with a  local  approximation  using   analytical methods for the most simple 
rotation law for constant specific angular momentum, i.e. the Rayleigh limit. Global calculations of the stability of the same rotation law  between two rotating perfect-conducting cylinders and threaded by a helical current-free  magnetic field lead to an almost perfect coincidence of the results of both methods. This is no longer true, however, for more flat rotation laws such as quasikeplerian rotation in a finite gap where the differences of the short-wave approximation  (which only holds for infinitely thin gaps) and global models are so strong that the results completely differ (R\"udiger \& Hollerbach 2007).

%%%%%%%%%%%%%%%%%%%%%%%%%%%%%%%%%%%%%%%%%%%%%%%%%%%%%%%%%%%%%%%%%%%%%%%%%%%%%%%%%%%%%%
\section{Dispersion relation for very small gaps}
%%%%%%%%%%%%%%%%%%%%%%%%%%%%%%%%%%%%%%%%%%%%%%%%%%%%%%%%%%%%%%%%%%%%%%%%%%%%%%%%%%%%%%
The dynamics of conducting fluids is described by the
 MHD equations
\begin{eqnarray}\label{1}
  \lefteqn{\frac{\p\bfu}{\p t}+(\bfu\cdot\nabla)\bfu=-\frac{1}{\rho}\nabla\left(p+
  \frac{\bfB^2}{2\mu_0}\right)+}\nonumber\\
&& \quad\quad\quad\quad\quad\quad\quad\quad\quad \quad  +\ \frac{1}{\mu_0\rho}(\bfB\cdot\nabla)\bfB+\nu\Delta^2\bfu,\\
  \label{2}
  \lefteqn{\frac{\p\bfB}{\p
  t}={\rm rot}\ (\bfu\times \bfB) \ +\ \eta\Delta^2\bfB,}
  \end{eqnarray}
and
${\rm div}\bfu= 
    {\rm div}\bfB=0.
$
Here $\bfu$ is the fluid velocity, $p$ the pressure, $\rho={\rm
const}$ the density and $\bfB$ is the magnetic field. In
cylindrical symmetry this set of equations 
has the stationary  solution 
$\bfB_0=B_{\phi}(R)\bfe_\phi+B_{z}\bfe_z$ where $B_{z}$ is a
constant and $B_{\phi}=I/R$  is current-free except on the rotation axis. 

Following Lakhin \& Velikhov (2007) we start with a local analysis. The
perturbations are  assumed to be
axisymmetric and proportional to $\exp(\gamma t+{\rm i}k_RR+{\rm i}k_zz)$,
where $\gamma=-{\rm i} \omega$ with (the real part of) $\omega$ as the eigenoscillation frequency  and
${\bfk=(k_R,0,k_z)}$  the wave number vector. The local short-wave  analysis is
justified when ${k_R\gg 1/R}$. Then the equations of
small-amplitude perturbations of the steady-state flow $\bfu_0= \Om(R) R$ are 
\begin{eqnarray}\label{8}
  \lefteqn{(\gamma+\omega_\nu) u_R'-2\Om u_\phi'=- {\rm i}k_R\frac{P'}{\rho}+
\frac{{\rm i}(\bfk\bfB_0)}{\mu_0\rho}B_R'\ -}\nonumber\\
&& \quad\quad\quad\quad\quad\quad\quad\quad\quad\quad\quad \quad \quad \quad -\ \frac{B_{0\phi}}
  {2\mu_0\rho R} B_\phi',\\
  \label{9}
\lefteqn{(\gamma+\omega_\nu) u_\phi'+\frac{\kappa^2}{2\Om} u_R'=
  \frac{{\rm i}(\bfk\bfB_0)}{\mu_0\rho} B_\phi',}\\
  \label{10}
  \lefteqn{(\gamma+\omega_\nu) u_z'=-{\rm i}k_z\frac{P'}{\rho}+\frac{{\rm i}(\bfk\bfB_0)}{\mu_0\rho} B_z',}\\
  \label{11}
  \lefteqn{(\gamma+\omega_\eta)B_R'={\rm i}(\bfk\bfB_0)u_R',}\\
  \label{12}
  \lefteqn{(\gamma+\omega_\eta) B_\phi'={\rm i}(\bfk\bfB_0) u_\phi'
  +R\frac{{\rm d}\Om}{{\rm d} R} B_R'+
 \frac{2B_{0\phi}}{R} u_R',}\\
  \label{13}
  \lefteqn{k_R u_R'+k_z u_z'=0, \quad\quad    k_R B_R'+k_z B_z'=0.}
  \label{14}
  \end{eqnarray}
Here $\omega_\nu=\nu k^2$ is the viscous frequency, 
$\omega_\eta=\eta k^2$ is the resistive frequency and the epicyclic frequency $\kappa$ is 
\begin{equation}
  \kappa^2=\frac{1}{R^3}\frac{{\rm d}}{{\rm d}R}\left(\Om^2R^4\right).
\end{equation}
$P'$ is the total pressure perturbation.
Then  the  dispersion relation
\begin{eqnarray}\label{16}
\lefteqn{  [(\gamma+\omega_\nu)(\gamma+\omega_\eta)+\omega_{\rm A}^2]^2
  + \frac{k_z^2}{k^2}\kappa^2\ \times}\nonumber\\
&& \quad \left[(\gamma+\omega_\eta)^2+\omega_{\rm A}^2+
  {\rm i}\frac{\omega_{\rm A}\omega_{{\rm A}\phi}}{\Om}(\omega_\nu-\omega_\eta)\right]-
    \phantom{\frac{k^2}{k_z^2}[(\gamma)]}\nonumber\\
\lefteqn{  -\ 4\frac{k_z^2}{k^2}
  [\Om\omega_{\rm A}+{\rm i}(\gamma+\omega_\eta)\omega_{{\rm A}\phi}]
 [\Om\omega_{\rm A}+{\rm i}(\gamma+\omega_\nu)\omega_{{\rm A}\phi}]}\nonumber\\
&& \quad\quad\quad\quad\quad\quad\quad\quad\quad\quad\quad\quad\quad\quad\quad\quad\quad\quad\quad =0
\end{eqnarray}
results with
\begin{equation}\label{17}
 \omega_{\rm A}^2=\frac{(\bfk\bfB_0)^2}{\mu_0\rho},\quad \quad\quad\quad\quad\quad \omega_{{\rm A}\phi}^2=
 \frac{B_{\phi}^2}{\mu_0\rho
 R^2}.
\end{equation}
The local dispersion relation is an equation of  fourth order
for the frequency.
In the two limiting cases when either $\omega_{{\rm A}\phi}=0$ or
$\omega_{\rm A}=0$ the coefficients in Eq.~(\ref{16}) are real. In  
general, however,   the coefficients of the dispersion 
relation (\ref{16}) are complex. To find  the stability criterion  
the method   by    Elstner, R\"udiger \& Tsch\"ape (1989) is applied. For ideal
fluids Blokland et al. (2005) used a very similar approach.

%%%%%%%%%%%%%%%%%%%%%%%%%%%%%%%%%%%%%%%%%%%%%%%%%%%%%%%%%%%%%%%%%%%%%%%%%%%%%%%%%%%%%%%%%%%%%%%%%%%%%%%
\section{Rotation with constant specific angular momentum (the Rayleigh limit)}\label{sec3}
%%%%%%%%%%%%%%%%%%%%%%%%%%%%%%%%%%%%%%%%%%%%%%%%%%%%%%%%%%%%%%%%%%%%%%%%%%%%%%%%%%%%%%%%%%%%%%%%%%%%%%
For rotation with constant specific angular momentum we have $\kappa=0$ which is called the Rayleigh limit in the Taylor-Couette community (see the MRI papers by  Willis \& Barenghi 2002; R\"udiger, Schultz \& Shalybkov 2003; Velikov et al. 2006). One obtains from Eq. (\ref{16})  the critical
angular velocity  
\begin{eqnarray}\label{vcrit}
\lefteqn{  \Om_c=\frac{k^2}{2k_z^2}\frac{F}{|\omega_{\rm A}|}\ \times}\nonumber\\
&&\left[\frac{
  (\omega_{\rm A}^2+\omega_\eta\omega_\nu)^2+4(k_z^2/k^2)\omega_\eta\omega_\nu
  \omega_{{\rm A}\phi}^2}{(\omega_\eta+\omega_\nu)^2\omega_{{\rm A}\phi}^2+
  (k^2/k_z^2)F(\omega_{\rm A}^2+\omega_\eta\omega_\nu)}\right]^{1/2},
\end{eqnarray}
with
$
  F\equiv\omega_{\rm A}^2+\omega_\eta\omega_\nu
  +2(k_z^2/k^2)\omega_{{\rm A}\phi}^2.
$
The frequency  of marginally stable modes is  
\begin{eqnarray}\label{vomega}
\lefteqn{  \omega=\omega_{{\rm A}\phi}\ \times}\nonumber\\
&&\left[\frac{
  (\omega_{\rm A}^2+\omega_\eta\omega_\nu)^2+4(k_z^2/k^2)\omega_\eta\omega_\nu
  \omega_{{\rm A}\phi}^2}{(\omega_\eta+\omega_\nu)^2\omega_{{\rm A}\phi}^2+
  (k^2/k_z^2)F(\omega_{\rm A}^2+\omega_\eta\omega_\nu)}\right]^{1/2}
\end{eqnarray}
(Lakhin 2007, private communication). For a very weak azimuthal magnetic
field
the instability limit is then defined by 
\begin{eqnarray}\label{vcrit1}
\lefteqn{  \Om_c^2=\frac{k^2}{4k_z^2}\frac{1}{\omega_{\rm A}^2}\ \times}\nonumber\\
\lefteqn{\left[\!
  (\omega_{\rm A}^2+\omega_\eta\omega_\nu)^2\!+\frac{k_z^2}{k^2}\omega_{{\rm A}\phi}^2
  \left[2(\omega_{\rm A}^2\!+\omega_\eta\omega_\nu)-(\omega_\eta-\omega_\nu)^2\right]\!\right]\!.}\nonumber\\
\end{eqnarray}
Obviously, when the condition 
\begin{equation}\label{resist}
  2(\omega_{\rm A}^2+\omega_\eta\omega_\nu)<
  (\omega_\eta-\omega_\nu)^2
\end{equation}
is fulfilled the  azimuthal magnetic field 
supports the instability. Hence, the destabilizing effect of
$B_\phi$ disappears for ${\rm Pm}=1$ but it also exists  for ${\rm Pm}>1$.

With the free parameter $\alpha=k_z/k$ and with 
\beg
\frac{\omega_{\rm A}}{\sqrt{\omega_\nu\omega_\eta}}= \alpha {\rm Ha}= {\rm Ha}^*,
\label{omA}
\ende
one finds from (\ref{vcrit}) and (\ref{vomega}) 
\begin{equation}
\frac{\omega}{\Om_c}=2\alpha^2 \frac{\omega_{{\rm A}\phi}\omega_{\rm A}}{F}
\label{omOm}
\end{equation}
and
$
F\simeq \omega_\nu \omega_\eta
$
for weak magnetic fields. Then
\begin{equation}
\frac{\omega}{\Om_c}= 2\alpha^2 \frac{\omega_{{\rm A}\phi}\omega_{\rm A}}{\omega_\nu \omega_\eta} \ll 1.
\label{omOm1}
\end{equation}
Hence, the oscillation frequency results  from the simultaneous 
existence of $\Om, B_z$ and $B_\phi$ (Knobloch 1996). It exists 
despite of  $\kappa=0$.  The magnetorotational instability  with 
helical but current-free magnetic fields is characterized by the 
existence of a frequency which only for the special case $B_\phi=0$  
vanishes (see also Blokland et al. 2005). As the simultaneous existence of 
poloidal and toroidal components of the magnetic field 
is quite characteristic for cosmic objects one could expect 
that the observation of the eigenfrequency may serve as the 
proof of the existence of the MRI (`magnetoseismology', see Blokland et al.). 

In the following the pitch angle
\begin{equation}
\beta=\frac{B_\phi}{B_z}
\label{BETA}
\end{equation}
is used, and we write 
$\beta^*=\beta/k_z R$
and
\beg 
{\rm Re}^*= \alpha {\rm Re}= \alpha \frac{\Om_c}{\omega_\nu}.
\label{alfRe}
\ende
Equations (\ref{vcrit}) and (\ref{vomega}) then lead  to
\begin{equation}
\frac{\omega}{\Om_c} {\rm Re^*} \simeq \alpha,
\label{omOmc}
\end{equation}
 meaning that
\begin{equation}
\omega\simeq\omega_\nu
\label{omomnu}
\end{equation}
for the marginal stable mode (Lakhin \& Velikhov 2007).
Obviously, even a small viscosity  proves to be  important for the system. Computations on the basis of the  dispersion relation (\ref{gamma}) under 
neglect of the viscosity (Liu et al. 2006) cannot lead  to the 
same conclusion. With the renormalization
\begin{equation}
\omega^*=\frac{\omega}{\sqrt{\omega_\nu \omega_\eta}},
\label{omstern1}
\end{equation}
one finds
\begin{equation}
\omega^*\simeq\sqrt{\rm Pm}\,,
\label{omstern2}
\end{equation}
independent of $\beta$.
At the Rayleigh limit it is thus  the viscosity frequency 
which is emanated by the MHD system. 

We shall show that this relation indeed is valid  
for the rotation law fulfilling the Rayleigh  
condition, i.e. ${\Om \propto R^{-2}}$. It remains 
true for both local and global calculations. 
Though, for more flat rotation 
profiles   (such as the Kepler law ${\Om \propto R^{-3/2}}$) the 
situation  becomes  more and  more  complicate,  
and the oscillation  frequency  starts to   run with 
the  Alfv\'en frequency $\omega_{\rm A}$ of the 
axial magnetic field (but not with the pitch angle $\beta^*$).  
Differences between the local and the global results become 
more and more large, only the  Pm-dependence of the characteristic 
frequency remains    weak.

%%%%%%%%%%%%%%%%%%%%%%%%%%%%%%%%%%%%%%%%%%%%%%%%%%%%%%%%%%%%%%%%%%%%%%%%%%%%%%%%%%%%%%%%%%%%%%%
\section{The numerical method}
\begin{figure}[t]%[htb]
\vskip -5mm \hskip -0.7cm
%    \center
    \includegraphics[width=8.5cm]{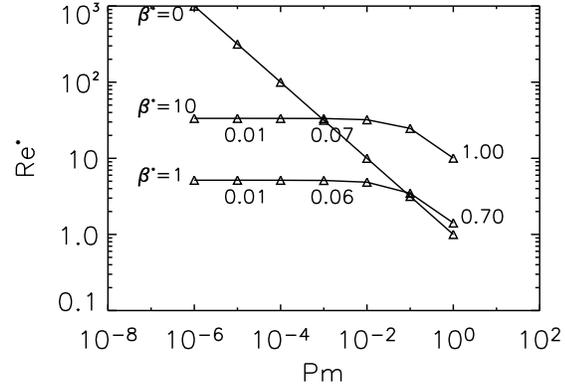}
\vskip -3mm
    \caption{Rayleigh limit: Critical Reynolds number vs. magnetic Prandtl number for axial fields ($\beta^*=0$) and for  helical fields ($\beta^*>0$). The curves are marked with the oscillation frequencies $\omega^*$ which obviously do not depend on the  $\beta$-values.}
    \label{fig1}
 \end{figure}

%%%%%%%%%%%%%%%%%%%%%%%%%%%%%%%%%%%%%%%%%%%%%%%%%%%%%%%%%%%%%%%%%%%%%%%%%%%%%%%%%%%%%%%%%%%%%%%%%%%%%%%%%%%
A numerical method to solve the dispersion relation has been developed. The dispersion relation  reads
\begin{equation}
\gamma^4+a_1\gamma^3+a_2\gamma^2+(a_3+{\rm i}b_3)\gamma+a_4+{\rm i}b_4=0,
\label{gamma}
\end{equation}
where
\begin{eqnarray}\label{24}
  \lefteqn{a_1=2(\omega_\eta+\omega_\nu),}\nonumber\\
  \lefteqn{a_2=(\omega_\eta+\omega_\nu)^2+2(\omega_{\rm A}^2
  +\omega_\eta\omega_\nu)+
\frac{k_z^2}{k^2}\kappa^2+
  4\frac{k_z^2}{k^2}\omega_{{\rm A}\phi}^2,}\nonumber\\
  \lefteqn{a_3=2(\omega_\eta+\omega_\nu)(\omega_{\rm A}^2
  +\omega_\eta\omega_\nu)
 +2\frac{k_z^2}{k^2}\kappa^2\omega_\eta\ 
+}\nonumber\\
&& \quad\quad\quad\quad\quad\quad\quad\quad \quad \quad \quad \quad +\   4\frac{k_z^2}{k^2}(\omega_\eta+\omega_\nu)\omega_{{\rm A}\phi}^2,\nonumber\\
  \lefteqn{a_4=(\omega_{\rm A}^2+\omega_\eta\omega_\nu)^2-4\frac{k_z^2}{k^2}\omega_{\rm A}^2\Om^2
  +\frac{k_z^2}{k^2}\kappa^2(\omega_{\rm A}^2+\omega_\eta^2)\ 
+}\nonumber\\
&& \quad\quad\quad\quad\quad\quad\quad\quad \quad \quad \quad \quad +\ 4\frac{k_z^2}{k^2}
  \omega_\eta\omega_\nu\omega_{{\rm A}\phi}^2,\nonumber\\
  \lefteqn{b_3=-8\frac{k_z^2}{k^2}\Om\omega_{\rm A}\omega_{{\rm A}\phi},}
  \nonumber\\
  \lefteqn{b_4=-4\frac{k_z^2}{k^2}\Om\omega_{\rm A}\omega_{{\rm A}\phi}(\omega_\eta+\omega_\nu)
  -\kappa^2\frac{k_z^2}{k^2}\frac{\omega_{\rm A}\omega_{{\rm A}\phi}}{\Om}
  (\omega_\eta-\omega_\nu)
.}\nonumber\\
\label{gamma1}
\end{eqnarray}
The dispersion relation (\ref{gamma}) is now solved numerically  for $\kappa=0$. To this end the (complex) frequency   is again normalized with $\sqrt{\omega_\nu\omega_\eta}$.
 Then from (\ref{24})
\begin{eqnarray}
\lefteqn{a_1= 2\left(\sqrt{\rm Pm}+ \frac{1}{\sqrt{\rm Pm}}\right),}\nonumber\\
\lefteqn{a_2= \left(\sqrt{\rm Pm}+ \frac{1}{\sqrt{\rm Pm}}\right)^2+ 2\left(1+{\rm Ha}^{*2}\right)+
 4\beta^{*2}{\rm Ha}^{*2},} \nonumber\\
\lefteqn{a_3=2\left(1+{\rm Ha}^{*2}\right)\left(\sqrt{\rm Pm}+ \frac{1}{\sqrt{\rm Pm}}\right)+}\nonumber\\
&& \quad\quad +\ 4\beta^{*2} {\rm Ha}^{*2}\left(\sqrt{\rm Pm}+ \frac{1}{\sqrt{\rm Pm}}\right),\nonumber\\
\lefteqn{a_4=\left(1+{\rm Ha}^{*2}\right)^2-4 {\rm Ha}^{*2} {\rm Re}^{*2} {\rm Pm} + 4\beta^{*2} {\rm Ha}^{*2} ,}\nonumber\\
\lefteqn{b_3= -8 \beta^* {\rm Ha}^{*2} {\rm Re^*} \sqrt{\rm Pm}\,,}\nonumber\\
\lefteqn{b_4=-4\beta^* {\rm Ha}^{*2} {\rm Re}^* (1+{\rm Pm}).}
\label{a1-4}
\end{eqnarray}
For given Pm, $\beta^*$ and ${\rm Ha}^*$ the dispersion relation  can  be solved in the following way. For too small ${\rm Re}^*$ all the solutions of (\ref{gamma}) will have negative $\Re(\gamma)$ so that the perturbations  decay. For a certain ${\rm Re}^*$, however, the first real part of one of the roots becomes positive so that the disturbance  grows and the flow becomes unstable. Then that ${\rm Ha}^*$ is searched which  provides the lowest ${\rm Re}^*$. This minimum ${\rm Re}^*$ is shown in Fig.~\ref{fig1} for various $\beta*$ in its dependence on the magnetic Prandtl number. Also the resulting normalized frequencies $\omega^*$ are given which indeed  do not depend on $\beta^*$. Note the basic difference of the solution for $\beta^*=0$ and $\beta^*>0$. Clearly for  ${\rm Pm}\to 0$ the solution with $\beta^*=0$ scales with ${\rm Rm}^*= {\rm Pm}\cdot {\rm Re}^*$ while the solutions with $\beta^*>0$ scale with Re. For ${\rm Pm}=1$ no basic difference exists between the solutions as there is no difference between Re and Rm. The consequence of the different scalings for small magnetic Prandtl number is a rather small critical ${\rm Re}^*$ for $\beta>0$.

One can also take from Fig.~\ref{fig1} that for $\beta^*=0$ ${\rm Re} \propto 1/\sqrt{\rm Pm}$. We know this relation as a particularity of the Rayleigh limit\footnote{It follows  directly from the relation $a_4=0$ for $\beta=\kappa=0$.} which changes to ${\rm Re}\propto 1/{\rm Pm}$ (i.e. ${\rm Rm}\simeq$ const) beyond the Rayleigh limit (see R\"udiger \& Hollerbach 2004). 

\begin{figure}[t]%[htb]
\vskip -5mm \hskip -0.8cm
%    \center
    \includegraphics[width=8.5cm]{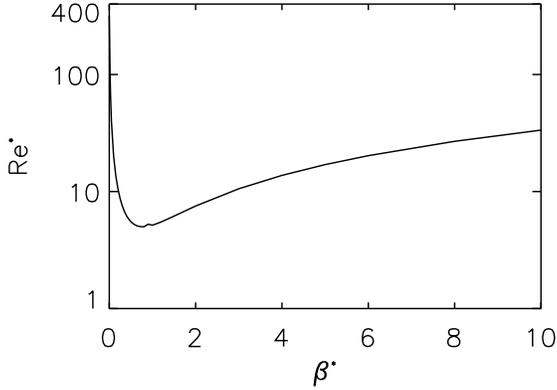}
\vskip -3mm
    \caption{Rayleigh limit: Critical Reynolds numbers vs. toroidal field component $\beta^*$ for fixed magnetic Prandtl number ($\rm Pm=10^{-5}$). On the vertical axis: $\rm Re^*=317, \ \rm Ha^*=1$.}
    \label{fig2}
 \end{figure}

We now ask  for the function ${\rm Re}^*={\rm Re}^*(\beta)$ for fixed and small Pm. Figure~\ref{fig2} shows  the  characteristic profile for  ${\rm Pm}=10^{-5}$. A minimum Reynolds number exists for $\beta^*\simeq 0.5$ which is smaller by two orders of magnitude than ${\rm Re}^*$ for $\beta^*=0$.

When the Reynolds number decreases for growing $\beta$ the frequency $\omega^*$ increases  to the value $\sqrt{{\rm Pm}}$ and remains then constant (see Eq.~\ref{omstern2}).  For small Pm it does not depend on $\beta$ and/or Re$^*$. The amplitude confirms the basic result (\ref{omomnu}).   

\begin{figure}[t]%[htb]
\vskip -5mm \hskip -0.8cm
%    \center
    \includegraphics[width=8.5cm]{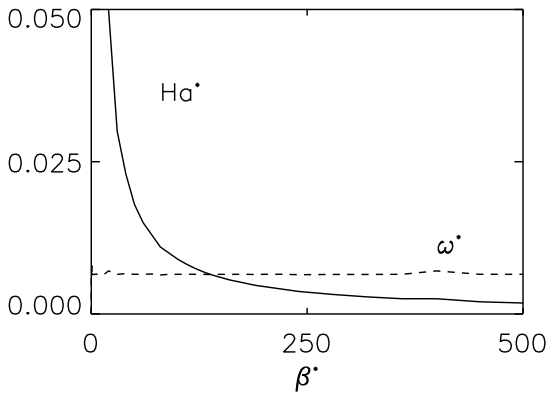}
\vskip -3mm
    \caption{Rayleigh limit: Hartmann number and oscillation frequency  for fixed magnetic Prandtl number ($\rm Pm=10^{-5}$)  for very strong toroidal field components.  Note the constancy of $\omega^*$ over the whole range of Reynolds numbers and  pitch angles $\beta^*>0$.}
    \label{fig2b}
 \end{figure}
Beyond the minimum  ${\rm Ha^*\propto 1/\beta^*}$ and ${\rm Re^*\propto \beta^*}$  which can be observed in Fig. \ref{fig2}.

In summary, the MRI for the Rayleigh limit $\Om \propto R^{-2}$ exists for not too high ratios  of azimuthal and axial   magnetic field components. It is basically oscillating  but not  for the exceptional case of  $B_\phi=0$.  The oscillation frequency approaches the viscous frequency $\omega_\nu$.   Pitch angels of order ten strongly destabilize the MRI, there is a deep and wide trough in the profile $\rm Re= Re({\beta})$ as shown in  Fig. \ref{fig2}.  Note that for  $\beta$ of order ten   the  scaling  with Re for $\rm Pm\to 0$ still exists which is  characteristic for HMRI.

%%%%%%%%%%%%%%%%%%%%%%%%%%%%%%%%%%%%%%%%%%%%%%%%%%%%%%%
\section{Kepler flow: local solutions}
\begin{figure}[t]%[htb]
\vskip -5mm
%    \center
    \includegraphics[width=8.5cm]{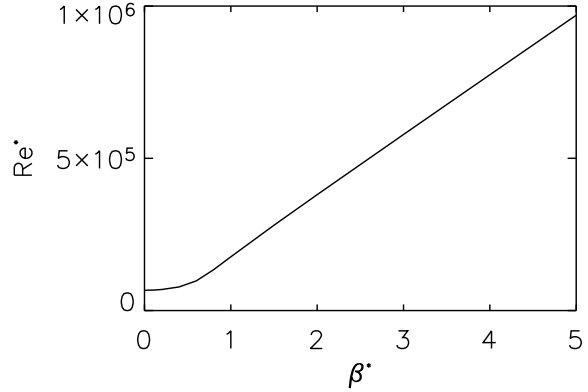}
\vskip -3mm
    \caption{ Kepler flow: The characteristic minimum in the Reynolds number profile  disappears. The toroidal field component stabilizes the MRI. On the vertical axis: $\rm Re^*= 66670, \rm Ha^*=259$. ${\rm Pm}=10^{-5}$.}
    \label{figkepler1}
 \end{figure}

%%%%%%%%%%%%%%%%%%%%%%%%%%%%%%%%%%%%%%%%%%%%%%%%%%%%%%%
%%%%%%%%%%%%%%%%%%%%%%%%%%%%%%%%%%%%%%%%%%%%%%%%%%%%%%%%%%%%%
In order to model the flat rotation law of a Kepler flow we have to work with  (\ref{24}) and $\kappa=\Om$. The numerical results are shown in Fig.~\ref{figkepler1}. They strongly differ from the results for the Rayleigh limit. The characteristic trough in the Reynolds number profile  in Fig.~\ref{fig2} disappears. The toroidal field  always stabilizes the MRI.
After Fig. \ref{figkepler1} the Reynolds number grows with $\beta^*$, i.e.
\beg
\rm Re^*\simeq 2 \cdot 10^5 \beta^*.
\label{rekep1}
\ende
 There is  no instability anymore with very small magnetic Reynolds number or small Lundquist number. 
 % (see Fig. \ref{figkepler3}).  
 Obviously, in the local approximation the solution in Fig. \ref{fig1} which scales for ${\rm Pm}\to 0$ with Re instead of Rm disappears  for too flat  rotation profiles.

\begin{figure}[t]%[htb]
\vskip -5mm \hskip -1cm
%    \center
    \includegraphics[width=8.5cm]{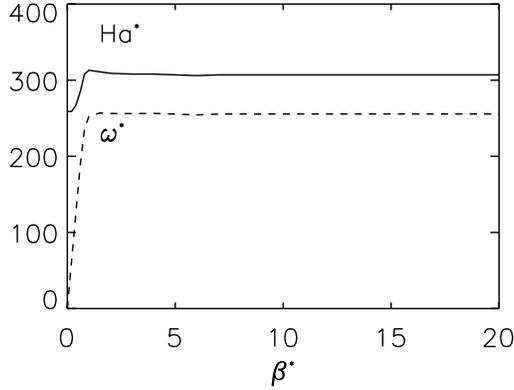}
\vskip -3mm
    \caption{Kepler flow: The Hartmann number $\rm Ha^* $ and the oscillation frequency $\omega^*$ are almost constant for $\beta^*>1$ and they  are almost equal so that again $\omega\simeq 0.8 \omega_{\rm A}$. ${\rm Pm}=10^{-5}$.}
    \label{figkepler2}
 \end{figure}

For the resulting frequency $\omega^*$  Fig. \ref{figkepler2} provides $\omega\simeq 0.8 \omega_{\rm A}$ independent of  the pitch angle $\beta$. In units of the rotation frequency, however,  
\beg
\frac{\omega}{\Om}\propto  \frac{1}{\beta^*}.
\label{rekep3}
\ende
This result   depends on the pitch angle of the magnetic field  but  does not depend on the magnetic Prandtl number. 
%%%%%%%%%%%%%%%%%%%%%%%%%%%%%%%%%%%%%%%%%%%%%%%%%%%%%%%%%%%%%%%%%%%%%%%%%%%%%%%%%%%%%%%%%%%%%%%%%%%%%%%%%%%%
\section{Global TC containers}
%%%%%%%%%%%%%%%%%%%%%%%%%%%%%%%%%%%%%%%%%%%%%%%%%%%%%%%%%%%%%%%%%%%%%
 With a local approximation we have considered the instability of differential rotation under the presence 
of a current-free magnetic field with helical geometry, i.e. $B_z B_\phi\neq 0$.  The local approximation only describes  flows with an extremely 
small radial extension. The instability  for such field geometries   
forms an overstable pattern of axisymmetric perturbations of flow and field. The corresponding 
eigenoscillation only vanishes for $B_\phi=0$. Beyond the Rayleigh limit Lakhin \& Velikhov (2007) find in the short-wave approximation that the eigenfrequency fulfills  $\omega\simeq \alpha \kappa$   without any  relation to  the magnetic field. 

The basically unknown quantity in local approximations is the resulting wave number $\alpha$. We shall switch, therefore, to the consideration of global models where the wave number no longer is an unknown quantity. We shall find that  the   eigenoscillation  $\omega$ of HMRI equals the  viscosity frequency also for global solutions for all pitch angels $\beta$ but only at the Rayleigh line. For more flat rotation laws the resulting eigenfrequency of the HMRI modes proves to be fixed by the Alfv\'en frequency of the vertical field.
\begin{figure}[t]%[htb]
\vskip -6mm \hskip -3mm
%    \center
    \vbox{
    \includegraphics[width=8.5cm]{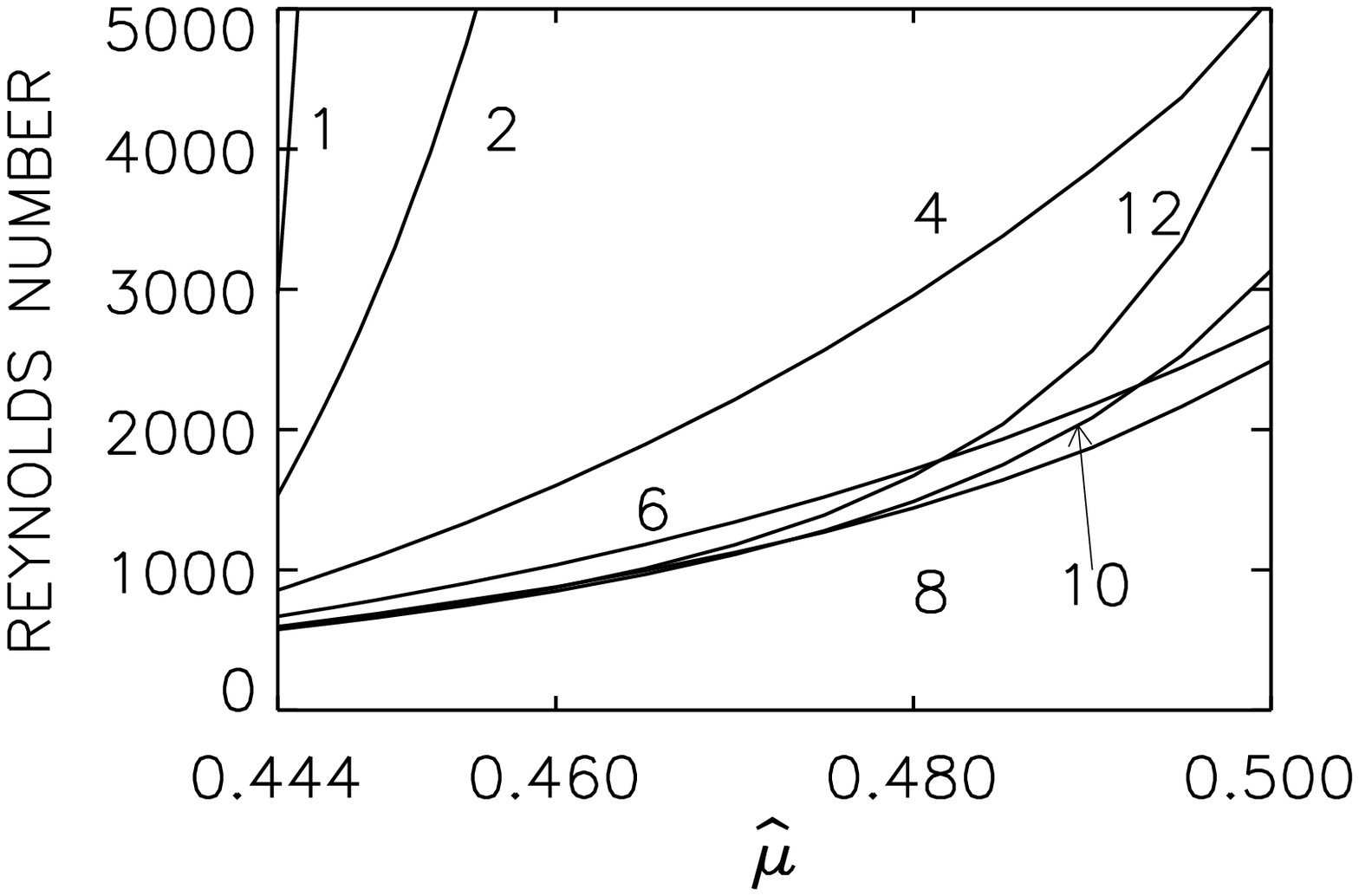} \\[-7mm]
     \includegraphics[width=8.5cm]{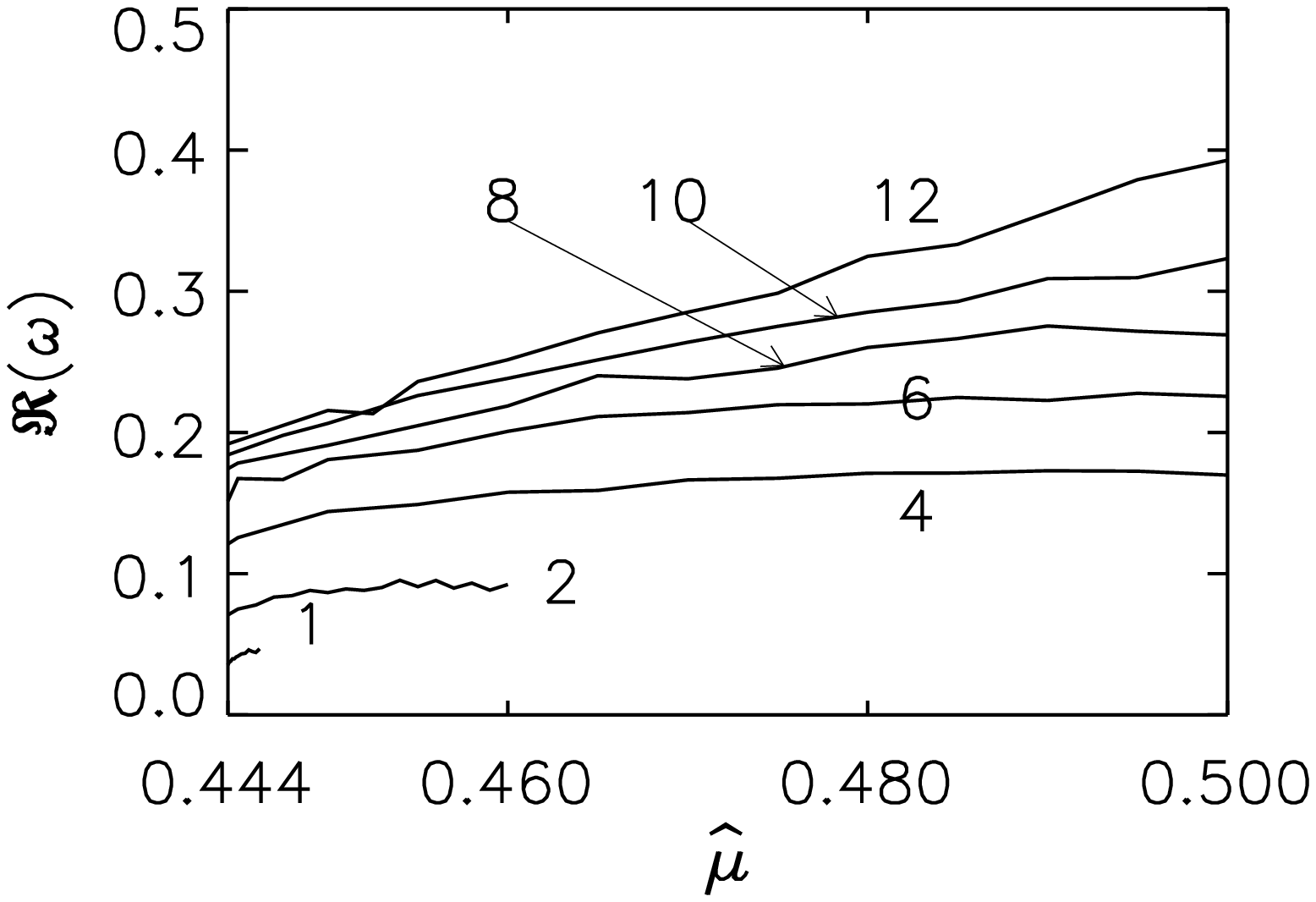}} \vskip -3mm
    \caption{Reynolds number (\emph{top}) and eigenfrequency (normalized with $\Om_{\rm in}$, 
    \emph{bottom}) for a small-gap container ($\hat\eta=0.67$) filled with gallium ($\rm Pm= 10^{-6}$) vs. the rotation rate ratio $\mu_{\Omega}$. The vertical axis at the left represents the Rayleigh limit. The curves are marked with  their $\beta$-values. Note the clear dependence of $\omega/\Om_{\rm in}$ on the pitch angle $\beta$.}
    \label{fig3a}
\end{figure}

For comparison with the results of the local approximation  we  first consider a  container with a small  gap 
between the cylinders ($R_{\rm in}/R_{\rm out}=0.67$) which are assumed as {\em perfect conductors}. The rotation law between the cylinders
is 
\begin{equation}
\Om= a+\frac{b}{R^2},
\label{om}
\end{equation}
with
 \begin{equation}
  a=\frac{{\mu_\Omega}-{\hat{\eta}}^2}{1-{\hat{\eta}}^2} \Om_{\rm in}, \ \ \ \ \ \ \ \ \ \ \ \ \ \  
  b=\frac{1-{\mu}_\Omega}{1-{\hat{\eta}}^2} R_{\rm in}^2 \Om_{\rm in},
 \label{ab}
 \end{equation}
 where
\begin{equation}
\hat{\eta}=\frac{R_{\rm in}}{R_{\rm out}}, \ \ \ \ \ \ \ \ \ \ \ \ \ \ \ 
\mu_\Omega=\frac{\Om_{\rm out}}{\Om_{\rm in}}.
\label{mu}
\end{equation}
$\Om_{\rm in}$ and $\Om_{\rm out}$ are the imposed rotation rates of
the inner and outer cylinders with their radii $R_{\rm in}$ and $R_{\rm out}$. The Rayleigh limit is reached if   $\mu_\Omega=\hat{\eta}^2$, i.e.  $\mu_\Omega=0.44$ for the considered  small-gap container. The axial field $B_z$ is uniform while the toroidal field is current-free in the  gap, i.e. $B_\phi/B_z=\beta R_{\rm in}/R$.The Hartmann number  and the
Reynolds number  are the dimensionless numbers of the problem,
\beg
{\rm{Ha}}=\frac{B_z R_0}{\sqrt{\mu_0 \rho \nu \eta}}, \ \ \ \ \ \ \ \ \ \ \ \ \ \ \ \ \ 
{\rm{Re}}=\frac{\Om_{\rm {in}} R_0^2}{\nu},
\label{pm}
\ende
where $R_0=(R_{\rm{in}}(R_{\rm{out}}-R_{\rm{in}}))^{1/2}$
is taken as the unit of length.  We have used $R_0^{-1}$ as the unit of
the wave number and 
$\Om_{\rm{in}}$ as the unit of frequencies.  We shall also
use the magnetic Reynolds number
$\rm Rm= Pm \cdot Re$. Details of the numerics can be found in R\"udiger et al. (2007a,b) and references therein.

\begin{figure}[t]%[htb]
\vskip -4mm \hskip -4mm
%    \center
    \vbox{
    \includegraphics[width=8.5cm]{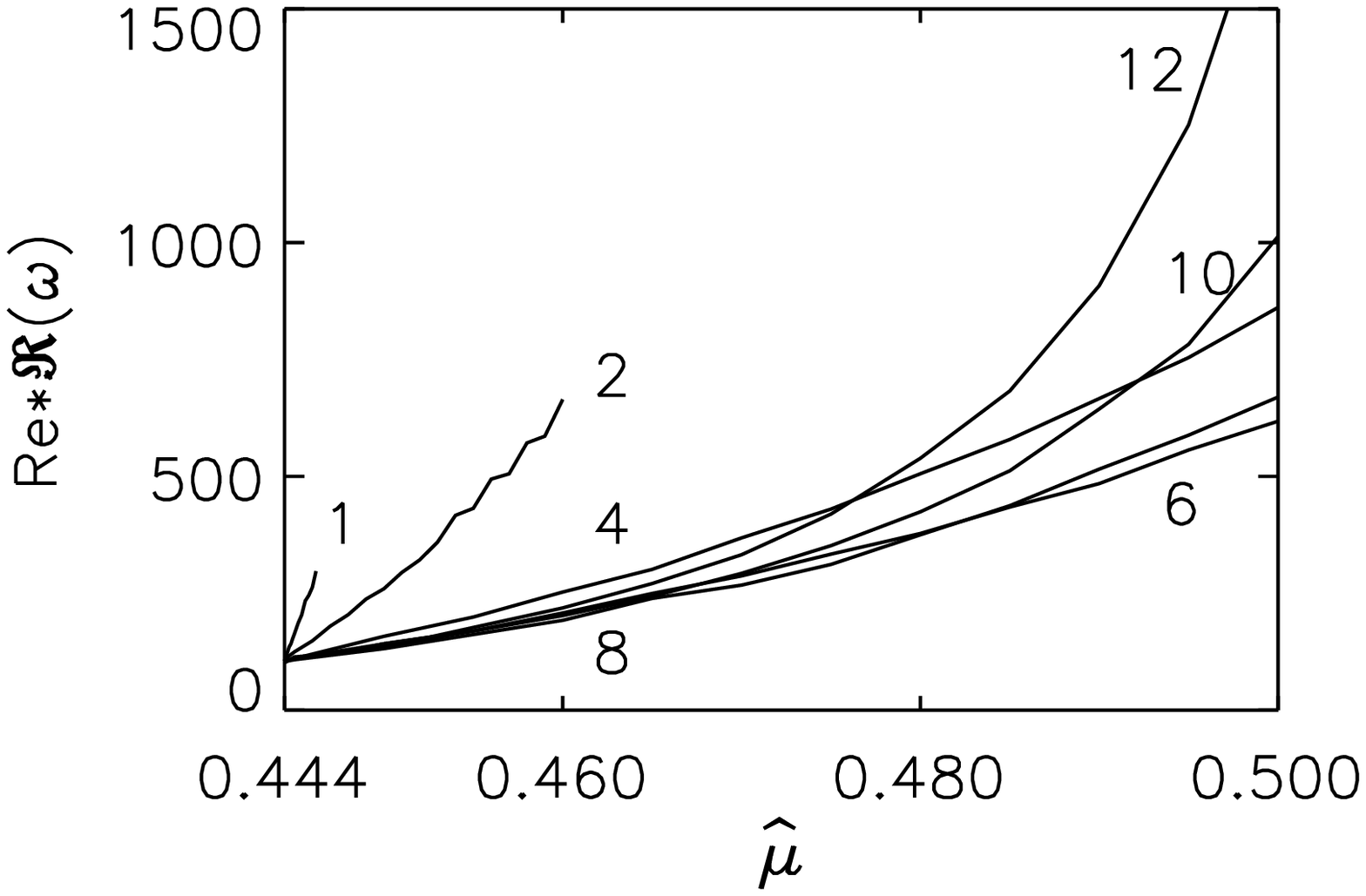} \\[-5mm]
    \includegraphics[width=8.5cm]{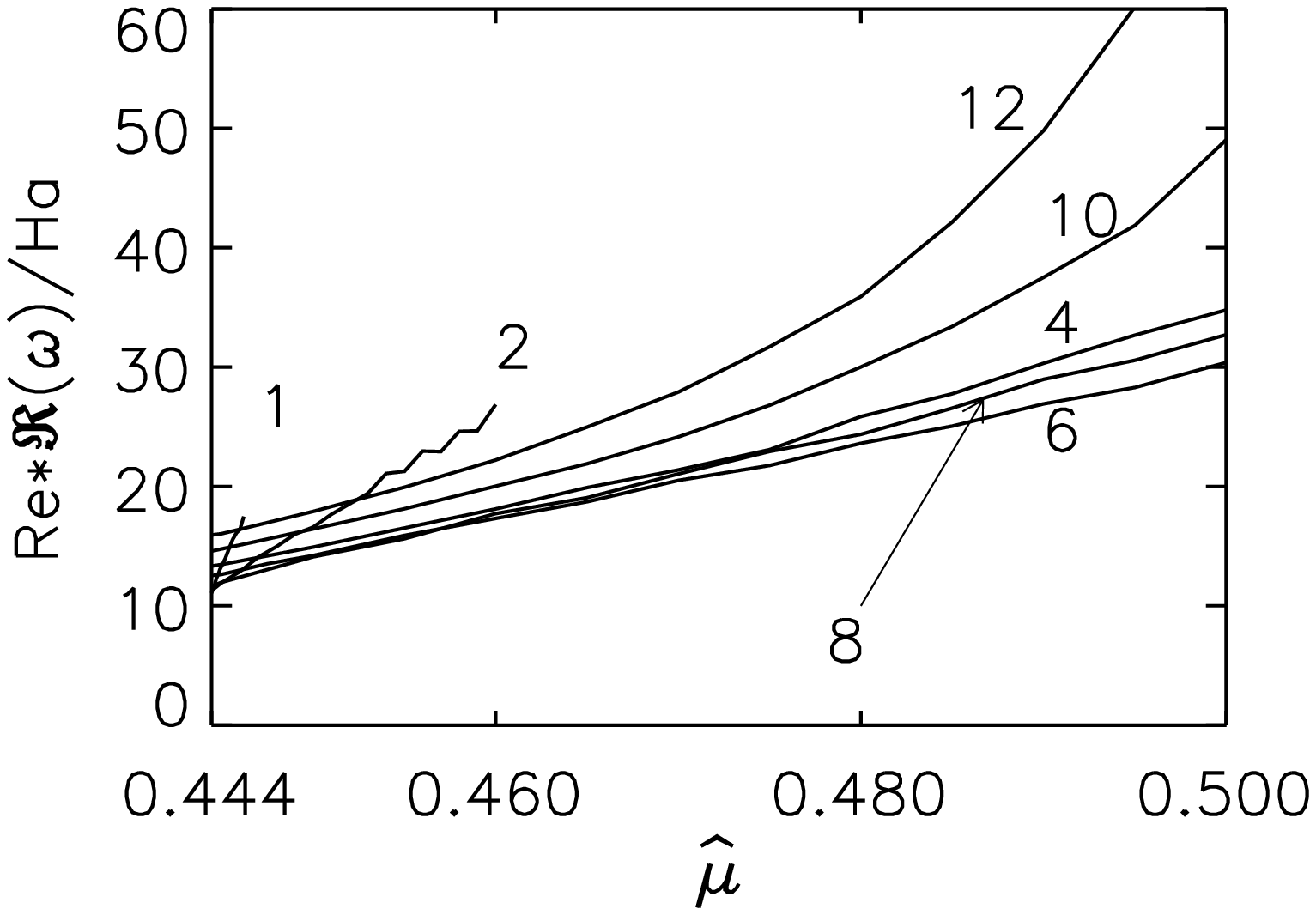}} \vskip -2mm
    \caption{The  same as in Fig. \ref{fig3a} but for the product of Reynolds number and  eigenfrequency 
  (\emph{top}) and this  product divided by Ha (\emph{bottom}). Note that at the Rayleigh line the product of Reynolds number and  eigenfrequency does not depend on $\beta$ and note that outside the Rayleigh line for medium $\beta$ the use of the magnetic normalization is successful.}
    \label{fig3b}
 \end{figure}

In Fig. \ref{fig3a} the Reynolds number (top) 
and the normalized oscillation frequency $\omega/\Om_{\rm in}$ (bottom) are given for very small magnetic Prandtl number ($\rm Pm = 10^{-6}$) and for various values of $\beta$. The main result is that for larger  $\beta$  the critical Reynolds decreases and  the normalized (!)
oscillation frequency grows. In consequence, the product of Reynolds number and oscillation frequency $\omega/\Om_{\rm in}$ should 
not vary to much with the $\beta$-values (see Fig. \ref{fig3b}, top). At the Rayleigh limit we indeed find one and 
the same value for the considered product independent of $\beta$  leading back  to  (\ref{omOmc}) yielding the basic relation (\ref{omomnu}). At the Rayleigh line the oscillation frequency is thus exactly given by the viscosity frequency. This is not true, however, beyond  the  Rayleigh line. There we have the  phenomenon that the oscillation frequency normalized with the viscosity frequency strongly depends on the $\beta$-value (Fig. \ref{fig3b}, top).  However, the $\beta$-dependence beyond the Rayleigh line is reduced if the Hartmann number of the axial field is used  (Fig. \ref{fig3b}, bottom).
For (medium) $\beta\simeq 4$--$8$ the curves show  ${\rm Re}\cdot (\omega/\Omega_{\rm in})\propto \rm Ha$ independent of $\beta$. This means
 \begin{equation}
\omega\propto  \omega_{\rm A},
\label{oma}
\end{equation}
almost independent of $\beta$  for medium values of $\beta$. A weak dependence on $\mu\Omega$ remains. Equation (\ref{oma}) means that for given magnetic Prandtl number the amplitude of the {\em vertical} field determines the oscillation frequency. This relation is not valid  for too small nor  too large values of $\beta$. For too small $\beta$ the influence of the toroidal field is too weak and for too large $\beta$ the influence of the axial field is too weak.

 %%%%%%%%%%%%%%%%%%%%%%%%%%%%%%%%%%%%%%%%%%%%%%%%%%%%%%%%%%%%%%%%%%%%%%%%%%%%%%%%%%%%%%%%%%%%%%%%%%
\section{Kepler flow: global solutions}
%%%%%%%%%%%%%%%%%%%%%%%%%%%%%%%%%%%%%%%%%%%%%%%%%%%%%%%%%%%%%%%%%%%%%%%%%%%%%%%%%%%%%%%%%%%%%%%%%%%%%%%%%%%
In contrast to the previous results we have demonstrated with nonlocal calculations that even for Kepler rotation HMRI occurs if  at least one of the boundaries is sufficiently conducting (R\"udiger \& Hollerbach 2007). The short-wave  method used in the foregoing sections does not reflect these results. In the following the Taylor-Couette calculations for flat rotation laws have been repeated i) to probe the basic results with another code and ii) to find more details  about the dependence of this effect on the magnetic Prandtl number and  about the eigenfrequencies and iii)  to include nonaxisymmetric modes  into the considerations. In order to obtain  exact solutions we remain in the frame of the Taylor-Couette flows so that only quasi-Kepler rotation laws can be considered. A quasi-Kepler rotation law  fulfills at both cylinders the Kepler law $\Om \propto R^{-1.5}$ so that the condition $\mu_\Omega=\hat{\eta}^{1.5}$ results. Here (as also in R\"udiger \& Hollerbach 2007)  with the standard gap of $\hat\eta=0.5$ is worked so that $\mu_\Omega=0.35$ holds for the quasi-Kepler approximation.

\begin{figure}[t]%[htb]
\vskip -0.8cm \hskip -6mm
%    \center
    \includegraphics[width=8.8cm, height=8cm]{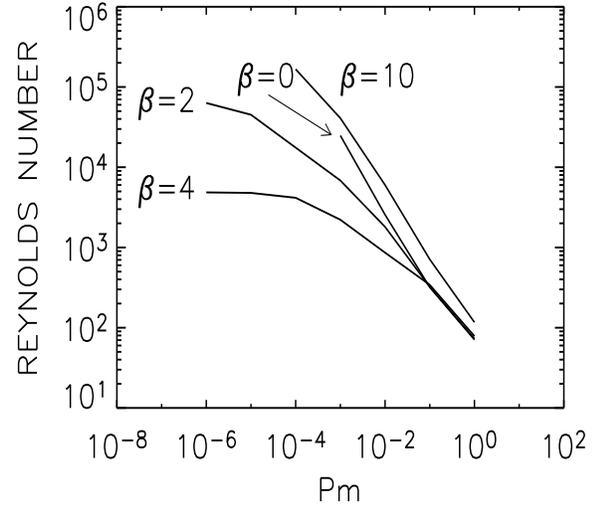} \vskip -5mm
    \caption{The critical Reynolds numbers for quasi-Kepler rotation.  Note that the fields with $\beta=0$ and $\beta=10$ for $\rm Pm\to 0$ scale with Rm and the fields with  $\beta=2$ and $\beta=4$ scale with Re. Perfect conducting cylinders, $m=0$, $\hat\eta=0.5$.}
    \label{fig5}
 \end{figure}
 
\begin{figure}[t]%[htb]
\vskip -5mm \hskip -4mm
%    \center
    \vbox{
    \includegraphics[width=8.8cm, height=6cm]{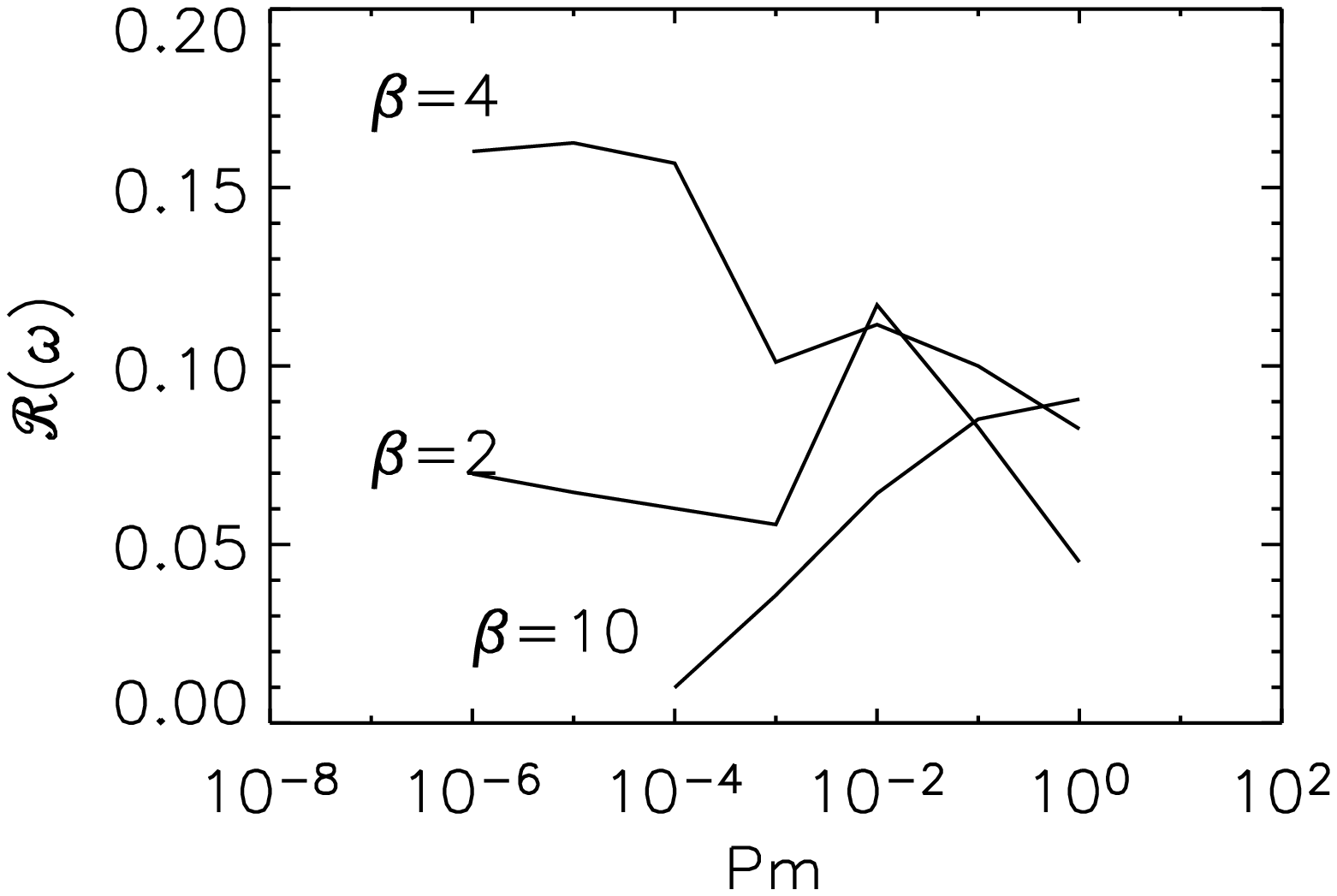} \\[-2mm]
     \includegraphics[width=8.8cm, height=6cm]{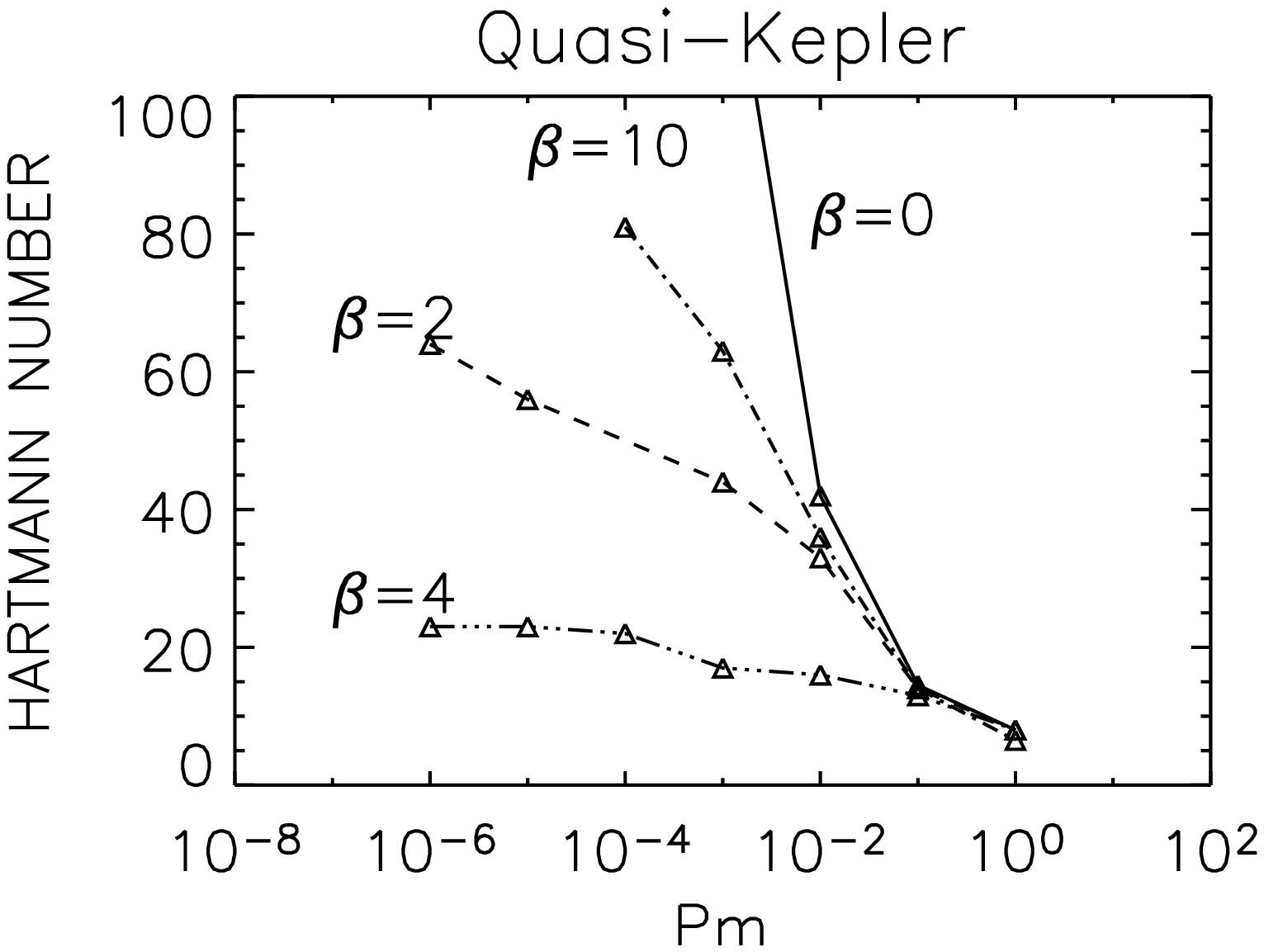} \\[-4mm]
      \includegraphics[width=8.5cm,height=5cm]{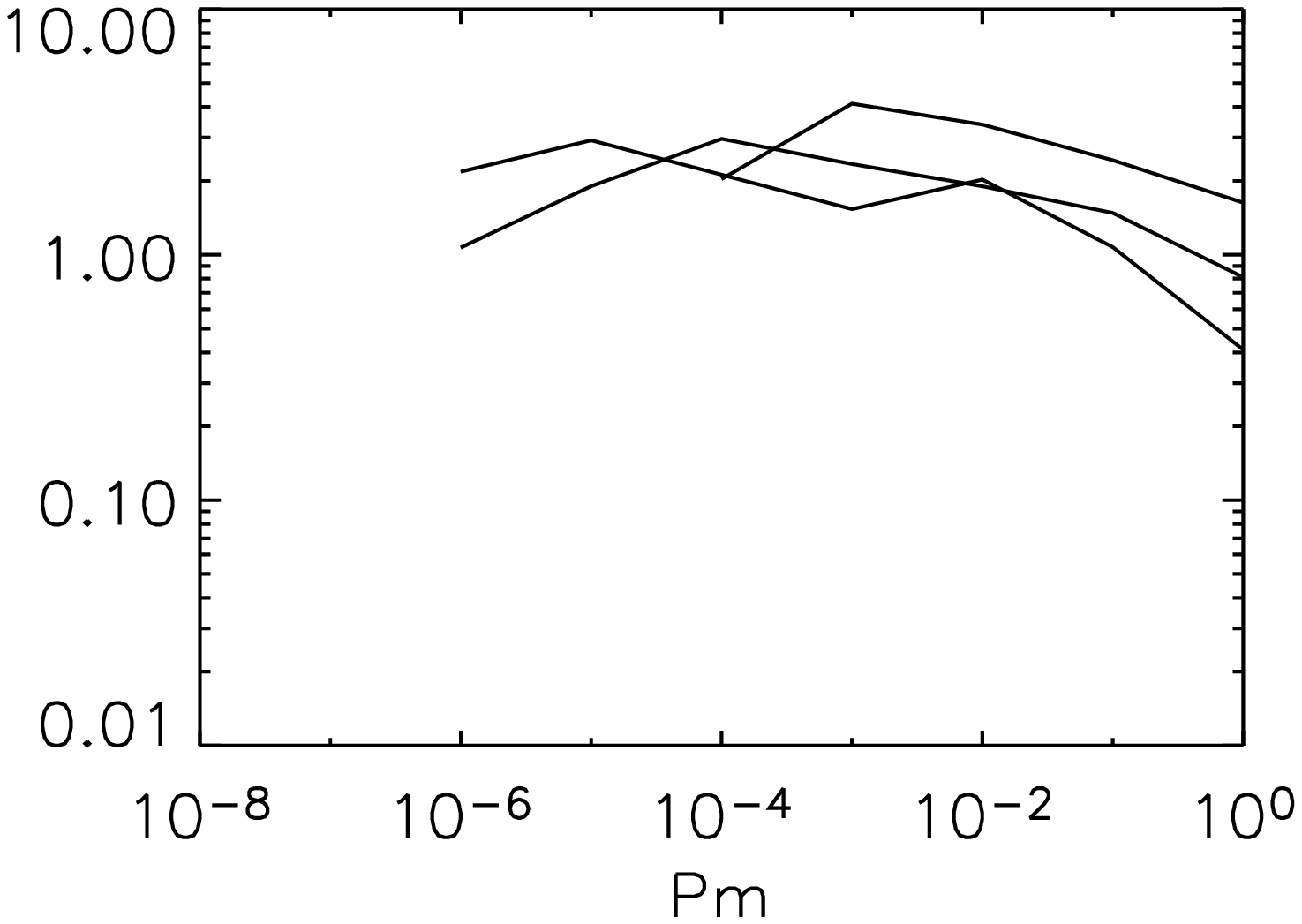}
     }
\vskip -3mm
    \caption{The normalized eigenfrequency ($\omega/\Om_{\rm in}$, \emph{top}), the  critical Hartmann  numbers (\emph{middle}) and the numerical basis of Eq. (\ref{ReHa}) for quasi-Kepler rotation (\emph{bottom}). 
In the bottom plot  the curves are for  $\beta=2, 4, 6$. Parameters as in Fig. \ref{fig5}.}
    \label{fig6}
 \end{figure}

The experiment PROMISE works with   a rotation law with $\mu_\Omega=0.27$. 
The experimental setup utilizes the existence of the trough in the relation of Re and $\beta$ where the critical  
Reynolds number of the  instability becomes very low. For more flat rotation laws one expects 
an increase of the  
critical Reynolds number. Lakhin \& Velikhov (2007) have shown  that  for $\kappa=\Om $ an instability 
   exists, with frequencies about one  order of magnitude smaller than the basic 
rotation, but the trough for medium $\beta$-values disappears. All the solutions are now  scaling 
with Rm rather than with Re. We shall show, however, that this result does {\em not} reflect the situation in Kepler disks which  obviously cannot be 
described in a local approximation.

Figure \ref{fig5} gives the critical Reynolds numbers as a function of $\beta$ and Pm.
It is clearly shown that for small $\beta$ ($\beta=0$) and large $\beta$ ($\beta=10$) the instability scales  with Rm while  for ${\rm Pm}\to 0$ the instability for $\beta=2$ and $\beta=4$ scales  with Re. This is the usual constellation for HMRI. For Kepler rotation there is no basic difference  close to the Rayleigh line. For ${\rm Pm}=1$ no distinction exists between Re and Rm  so that  all  differences disappear  between the solutions for  different  $\beta$. The oscillation frequencies $\omega/\Om_{\rm in}$ are   of order 0.1 (with a weak $\beta$-dependence for small Pm) and they  only slightly depend on the magnetic  Prandtl number (see Fig.~\ref{fig6}, top). One finds
\beg
\frac{\omega}{\Om_{\rm in}}\frac{{\rm Re}}{{\rm Ha}} {\rm Pm}^{1/4}\simeq 1
\label{ReHa}
\ende
for  Pm between  $10^{-6}$ and 1 (Fig.~\ref{fig6}, bottom). Hence
\beg
\omega= \omega_{\rm A}\ {\rm Pm}^{1/4},
\label{omsim}
\ende
independent of $\beta$. Only the amplitude of the axial field fixes  the eigenfrequency. The influence of the magnetic Prandtl number is small. This is a numerical result for not too small Pm, the physical meaning of the Pm-factor is not yet completely  clear. The result (\ref{omsim}) for  ${\rm Pm}=10^{-5}$ is smaller by  one order of magnitude than the result  derived  from  the  dispersion relation.

%%%%%%%%%%%%%%%%%%%%%%%%%%%%%%%%%%%%%%%%%%%%%%%%%%%%%%%%%%%%%%%%%%%%%%%%%%%%%%%%%%%%%%%%%%%%%%%%%%%%%%%
\subsection{Nonaxisymmetric modes}
%%%%%%%%%%%%%%%%%%%%%%%%%%%%%%%%%%%%%%%%%%%%%%%%%%%%%%%%%%%%%%%%%%%%%%%%
Also  nonaxisymmetric solutions  with $\exp({\rm im\phi})$  have been computed with this model which  represent solutions drifting 
along the azimuth. R\"udiger et al. (2005) find the solution of this problem for $m=1$ for various $\mu_\Omega$ but only for a fixed 
magnetic Prandtl number. Always the mode with $m=1$ needs a higher Reynolds number to be excited than the mode with $m=0$. At 
the Rayleigh limit and for $\rm Pm=10^{-5}$ the critical Reynolds number proves to be 3$\cdot 10^6$. 

\subsubsection{Standard MRI}
\begin{figure}[t]%[htb]
\vskip -7mm \hskip -5mm
%    \center
    \vbox{
    \includegraphics[width=8.8cm, height=6cm]{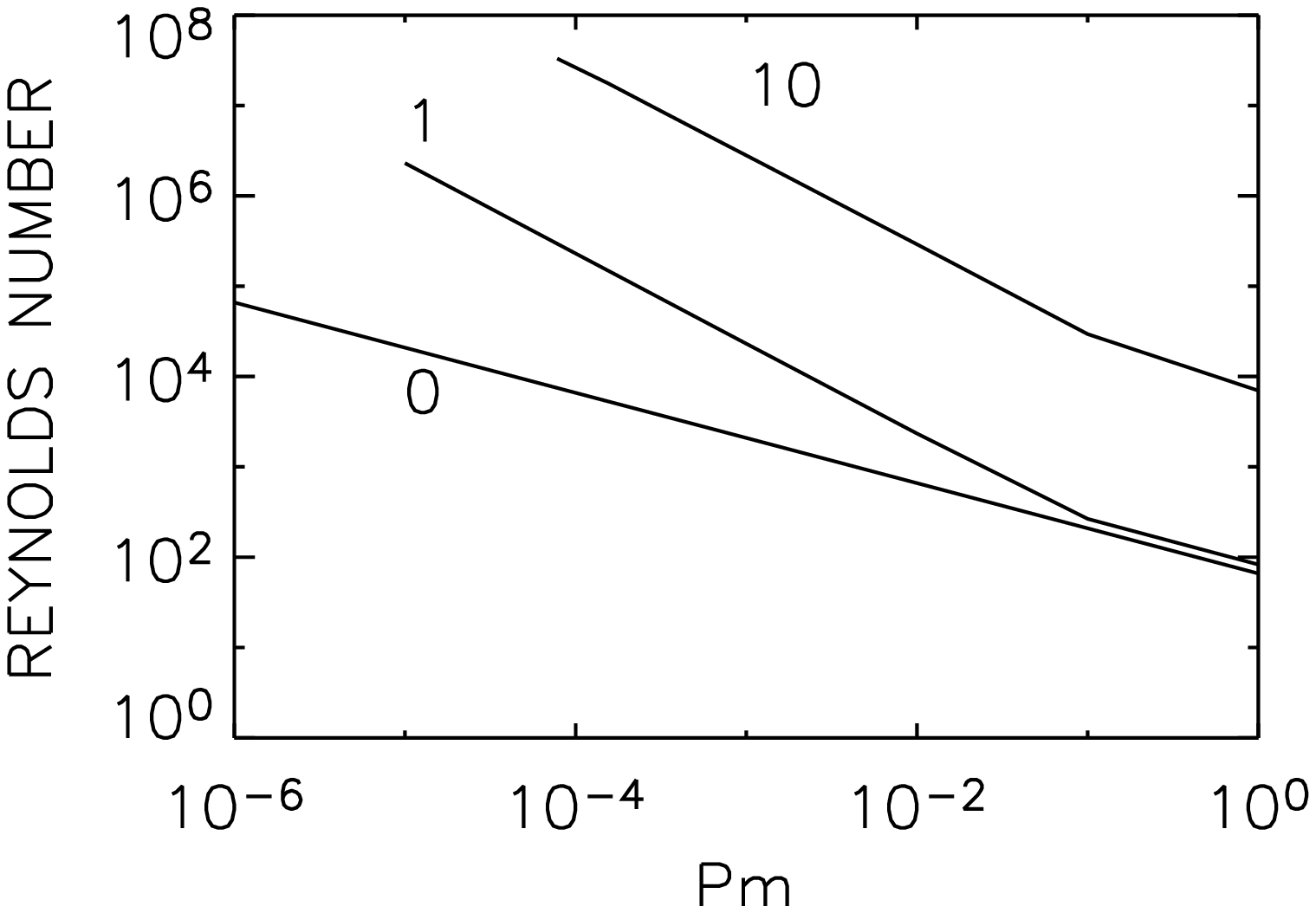}  
    \includegraphics[width=8.8cm, height=6cm]{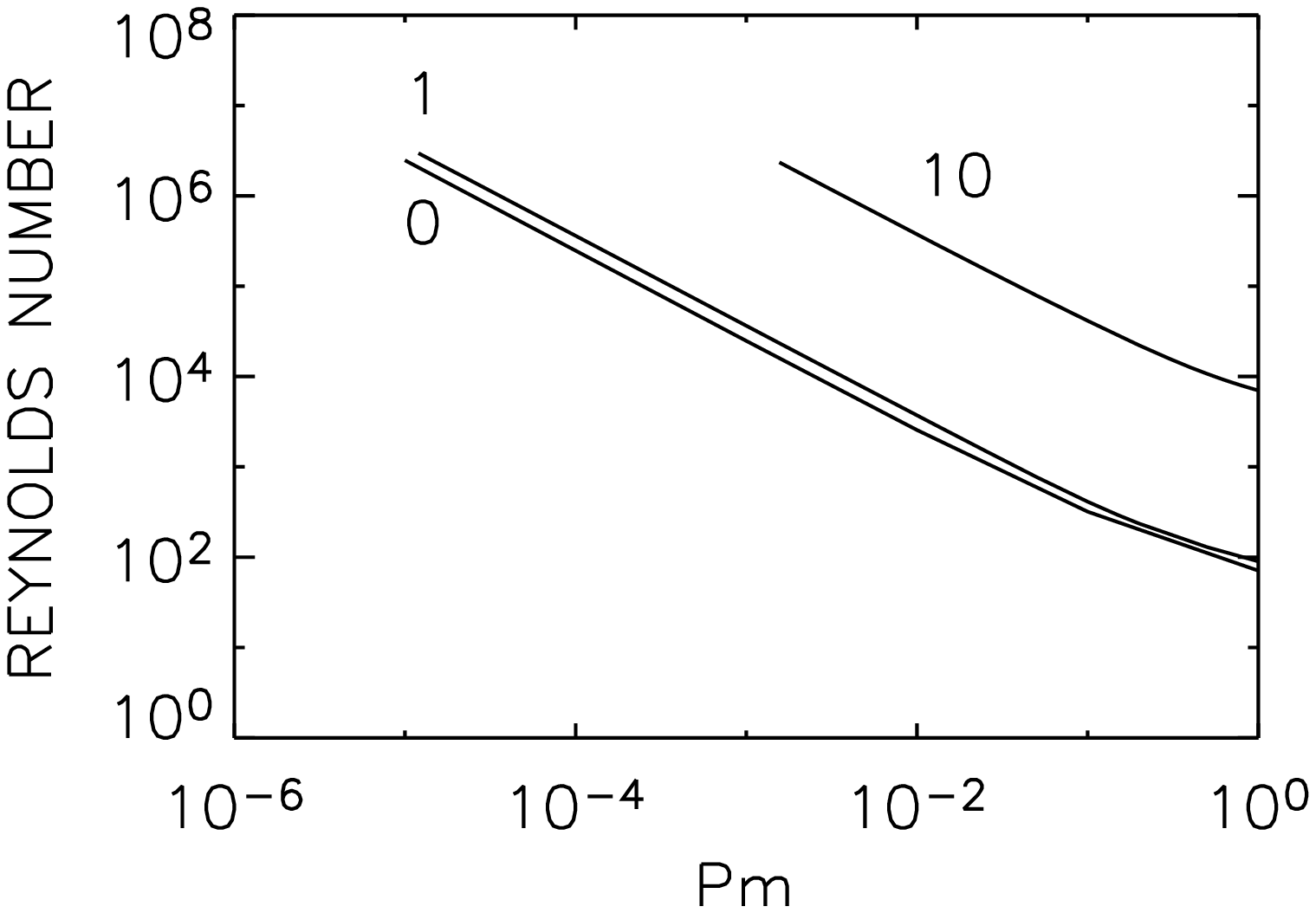} }
    \caption{The critical Reynolds numbers for standard MRI (i.e. $\beta=0$) vs. magnetic Prandtl number at the Rayleigh limit ({\em top}) and for quasi-Kepler
    rotation ({\em bottom}). The curves are marked with their mode number $m$. In both cases for small Pm the differences are huge.}
    \label{fig15}
 \end{figure}

%%%%%%%%%%%%%%%%%%%%%%%%%%%%%%%%%%%%%%%%%%%%%%%%%%%%%%%%%%%%%%%%%%%%%%%%
For standard MRI  (i.e. $\beta=0$, see R\"udiger \& Zhang 2001; Ji, Goodman \& Kageyama 2001) the calculations have been repeated for various 
magnetic Prandtl numbers and for $m=0 , 1, 10$ (Fig. \ref{fig15}). For all nonvanishing $m$ the critical Reynolds numbers  lie {\em above} the curve for $m=0$. It is 
also understandable that this effect grows for smaller Prandtl numbers. The differential rotation increases the magnetic dissipation of nonaxisymmetric 
patterns growing  with $m^2 \eta$ and $\eta$ runs with $1/\rm Pm$. This situation exists for both steep rotation law (top) and also for  flat rotation laws 
such as the Keplerian one (bottom). In
general, the nonaxisymmetric modes are strongly stabilized, by the smoothing action of differential rotation 
but also by the appearance of the azimuthal drift as argued by  Mikhailowskii et al. (2008). Standard MRI is obviously a basically axisymmetric phenomenon.

 %%%%%%%%%%%%%%%%%%%%%%%%%%%%%%%%%%%%%%%%%%%%%%%%%%%%%%%%%%%%%%%%%%%%%%%%%%%%%%%%%%%%%%%%%%%
\subsubsection{HMRI}
\begin{figure}[t]%[htb]
\vskip -7mm  \hskip -5mm
%    \center
    \includegraphics[width=8.8cm, height=6.5cm]{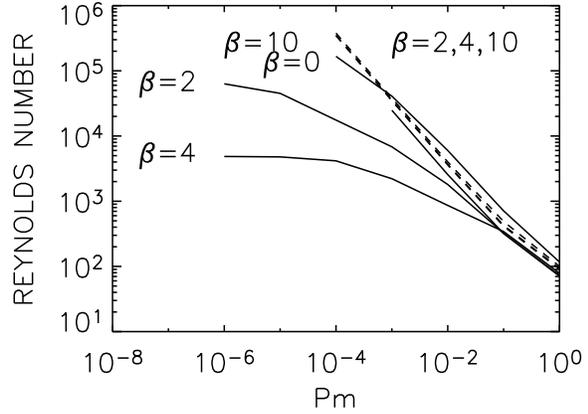}  \vskip -5mm
    \caption{The same as in Fig. \ref{fig5} but with the nonaxisymmetric mode $m=1$ included. The solid lines give the axisymmetric 
    modes ($m=0$) for various $\beta$ while the dashed line belongs to the nonaxisymmetric ($m=1$) modes for all $\beta$.}
    \label{fig9}
 \end{figure}

The  motivation for the inclusion of nonaxisymmetric modes for helical magnetic geometry is   that for large $\beta$ the toroidal field  
dominates the axial field so that previous computations concerning the $m=1$ instability for azimuthal (current-free) magnetic fields 
with $B_\phi \propto 1/R$ (AMRI) should be concerned. In Fig.~\ref{fig9} the dashed line gives the instability limit for $\beta=10$. One 
finds that it scales with Rm for  ${\rm Pm}\to 0$ also known for AMRI (see R\"udiger et al. 2007b). 
 \begin{figure}[t]%[htb]
\vskip -0.8cm   \hskip -4mm
%    \center
    \includegraphics[width=8.8cm, height=6.5cm]{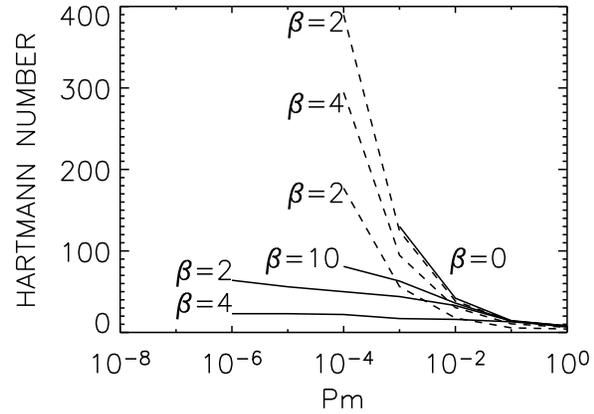}  \vskip -5mm
    \caption{The same as in Fig. \ref{fig9} but for the Hartmann number.}
    \label{fig10}
 \end{figure}

For the mode with $m=1$ the critical Reynolds number for small Pm exceeds the critical Reynolds number for $m=0$ so that the 
axisymmetric traveling-mode is {\em much easier} to excite than the nonaxisymmetric solution. The opposite is true for ${\rm Pm}\rsim 10^{-3}$. In this 
case, for $\beta \geq 10$, the nonaxisymmetric modes  are more  easy to excite.

We also computed the nonaxisymmetric solutions  for $\beta=2$ and $\beta=4$.  We find that their instability limits can not be 
distinguished from that for $\beta=10$. It is clear that also  the solutions with  $\beta>10$ comply with the curve for $\beta=10$. 
The Hartmann number, however, becomes smaller and smaller for growing $\beta$ which, however, is not a surprise as for azimuthal 
MRI (also for  the Tayler instability) the toroidal field strength is important, i.e. the product $\beta \ {\rm Ha}$.

Note that for  small Pm  the Hartmann numbers for $m=1$ are {\em much} higher than for $m=0$ (Fig.~\ref{fig10}). This may be a simple 
consequence of the fact that AMRI scales with the Lundquist number ${\rm S}=\sqrt{{\rm Pm}}\ \beta \ {\rm Ha}$.
%%%%%%%%%%%%%%%%%%%%%%%%%%%%%%%%%%%%%%%%%%%%%%%%%%%%%%%%%%%%%%%%%%%%%%%%%%%%%%%%%%%%%%%%%%%%%%%%%%%
\subsection{Results}
%%%%%%%%%%%%%%%%%%%%%%%%%%%%%%%%%%%%%%%%%%%%%%%%%%%%%%%%%%%%%%%%%%%%%%%%%%%%%%%%%%%%%%%%%%%%%%%%%%%%%%
The basic  results from our calculations for quasi-Keplerian rotation laws are: 
\begin{itemize}

\item[1.]
For $0<\beta\lsim 4$  and $\rm Pm<1$  the axisymmetric HMRI has always the lowest Reynolds number. The traveling-wave solution with $m=0$ which is observed in the 
PROMISE experiment  forms the instability  which is most easiest to excite also for Kepler rotation. Opposite statements (Liu et al. 2006) cannot be  confirmed.
\item[2.] The oscillation frequency for $\beta>0$ equals  the viscosity  frequency  but  only  at  the Rayleigh  line.  For  quasi-Kepler  rotation  it proves  
to be the Alfv\'en frequency  of the vertical magnetic  field.
\item[3.]
For larger $\beta$ (${\beta \simeq 10}$) and   ${\rm Pm \lsim 10^{-3}}$ the lowest Reynolds numbers belong to the axisymmetric HMRI 
while  for $\rm Pm > 10^{-3}$ they always belong to the nonaxisymmetric AMRI.
\item[4.]  The neutral instability line for $m=1$ does  not depend on $\beta$, i.e. the AMRI is not concerned by the existence of  axial fields.
\end{itemize}
The calculations demonstrate the strong  influence of the magnetic Prandtl number for the MHD instability theory. 
The formula 
SMRI (axisymmetric, stationary, scaling with Rm) + AMRI (nonaxisymmetric, drifting, 
scaling with Rm) = HMRI (axisymmetric, oscillating, scaling with Re)  exists only for small Pm.
%%%%%%%%%%%%%%%%%%%%%%%%%%%%%%%%%%%%%%%%%%%%%%%%%%%%%%%%

\end{document}